# Computational assessment of non-polar and polar GaP terminations for photoelectrochemical water splitting

*Sofia Apergi, Sreejith Pallikkara Chandrasekharan, Charles Cornet\*, Laurent Pedesseau\**

Univ Rennes, INSA Rennes, CNRS, Institut FOTON – UMR 6082, F-35000 Rennes, France

ABSTRACT

With photoelectrochemical water splitting being one of the most promising approaches for clean energy production and storage, the search for efficient photoelectrode materials is greater than ever. Gallium phosphide (GaP) is a well-established semiconductor with suitable band edge positions that has already been successfully employed in photoelectrochemical solar cells. However, to utilize it as efficiently as possible, a proper understanding of its properties when interfaced with water is required, and this is currently lacking. In this work we use ab initio molecular dynamics simulations to study the properties of the aqueous interfaces of various GaP non-polar (110) and polar (001) terminations. We calculate their band alignment with respect to the standard hydrogen electrode potential and investigate their interfacial structural properties.

Based on these properties we assess the capability of the various terminations to catalyze the reactions associated with water splitting and propose approaches for improving the performance of GaP for application in PECs.

INTRODUCTION

Photoelectrochemical water splitting can play a pivotal role in combating climate change, being one of the most promising approaches for sustainable energy production and storage[1,2]. This process can take place in photoelectrochemical cells (PECs), where photogenerated charge carriers drive the redox reactions that split water into $H_2$ and $O_2$[3,4]. These cells employ semiconducting photoelectrodes, which must possess an appropriate bandgap for optimal absorption of sunlight, as well as band edges that are properly aligned with the redox potentials corresponding to the Hydrogen Evolution Reaction (HER) and Oxygen Evolution Reaction (OER)[4]. Although numerous candidate photoelectrode materials are being explored for this purpose, the development of PECs that are simultaneously efficient, photochemically stable, and cost-effective remains a major challenge.

III-V semiconductors comprise a versatile class of materials, with excellent, highly tunable, optoelectronic properties, which make them suitable for application in PECs[4,5]. Among them, Gallium Phosphide (GaP) stands out, due to its potential to simultaneously promote both HER and OER. Due to the favorable position of its conduction band minimum (CBM), which lies approximately 1-1.5 eV above the standard hydrogen electrode (SHE) potential[6], GaP has been employed extensively as a photocathode for HER, with its first use in a PEC dating back to 1977[7]. Moreover, with a bandgap of ~2.3 eV, its valence band maximum (VBM) is then positioned very close to the water oxidation potential (1.23 eV below SHE), which enables GaP to catalyze the oxidation of water, as has also been proven exprimentally[8,9]. An additional advantage of GaP is its small lattice mismatch with silicon, which allows for high quality crystals to be grown epitaxially, thus offering a viable route towards device scalability[10].

Despite these promising properties, the widely cited band edge positions of GaP are typically derived from bulk calculations and often neglect key interfacial effects, such as atomistic details of the surface morphology, termination, and interaction with the electrolyte. These factors can have a huge impact on the electronic properties of the semiconductor[11–13], as well as on the charge transfer processes between the semiconductor and the electrolyte [6,14,15]. For example, they can shift the band edge positions of the former by more than 1 eV, as has been demonstrated by Guo et al. for the case of (110)-terminated GaP and Gallium Arsenide (GaAs), due to variations in water adsorption configurations at the interface[16]. In addition, favorable optoelectronic properties alone do not guarantee device efficiency, since other properties, such as the atomic-scale arrangement of the interface can also influence reaction kinetics[17]. Characterizing the structural and electronic properties of aqueous semiconductor interfaces is therefore crucial for realizing stable and efficient PECs and taking advantage of the full potential of the employed materials, especially considering that some fabrication techniques allow for very precise control of the surface morphology of the deposited semiconductor.

In this work we use ab initio molecular dynamics (AIMD) simulations to study the aqueous interfaces of various non-polar (110) and polar (001) GaP terminations. While the former have the advantage of not being reconstructed at the atomic level and are therefore simpler to describe theoretically, (110) terminations are rarely observed and used practically in III-V semiconductor applications, including PEC devices. The (001) facets, on the other hand, are the most commonly used surfaces in III-V devices and typically correspond to the top surface of standard commercial

III-V wafers. First, we determine the alignment of the semiconductor electronic levels with respect to SHE and explain these results based on the interfacial charge distribution and electric dipole arrangement. Specifically, we show how the semiconductor surface terminations can impact the electron transfer between the semiconductor surface and the electrolyte and, as a result, the electronic level alignment. In addition, by analyzing the structural properties of the studied interfaces we argue that favorable electronic level alignment alone does not necessarily translate into efficient photoelectrodes. Finally, we discuss the broader implications of our results in the context of PEC design and propose potential strategies for effective surface passivation.

RESULTS AND DISCUSSION

First, the position of the semiconductor band edges, namely the valence band maximum $\varepsilon_V^{SHE}$ and the conduction band minimum $\varepsilon_C^{SHE}$, relative to the redox potentials associated with HER and OER, at the GaP-water interface, are to be determined. This requires knowledge of the standard hydrogen electrode potential ($\mu_{SHE}$) and the bulk semiconductor band edges ($\varepsilon_V^b$ and $\varepsilon_C^b$). Once these quantities are known, $\varepsilon_V^{SHE}$ and $\varepsilon_C^{SHE}$ can be calculated by aligning the average electrostatic potential across the GaP-water interface[16,18–20], a process that is schematically described in Fig. 1. $\varepsilon_V^{SHE}$ is then given as:

$$\varepsilon_V^{SHE} = \varepsilon_V^b - \mu_{SHE} - \Delta V_{int}, \quad (1)$$

where $\Delta V_{int}$ is the electrostatic potential difference between the bulk-like regions of GaP and water in the solid/liquid heterostructure. Once $\varepsilon_V^{SHE}$ is known, $\varepsilon_C^{SHE}$ can be calculated as:

$$\varepsilon_C^{SHE} = \varepsilon_V^{SHE} + E_g \quad (2)$$

where $E_g$ is the bandgap of the semiconductor. It should be noted that $\mu_{SHE}$ here is referenced to the planar averaged electrostatic potential of bulk liquid water ($V_{\text{water}}$) and similarly $\varepsilon_V^b$ and $\varepsilon_C^b$ are referenced to the planar averaged electrostatic potential of bulk GaP ($V_{\text{GaP}}$).

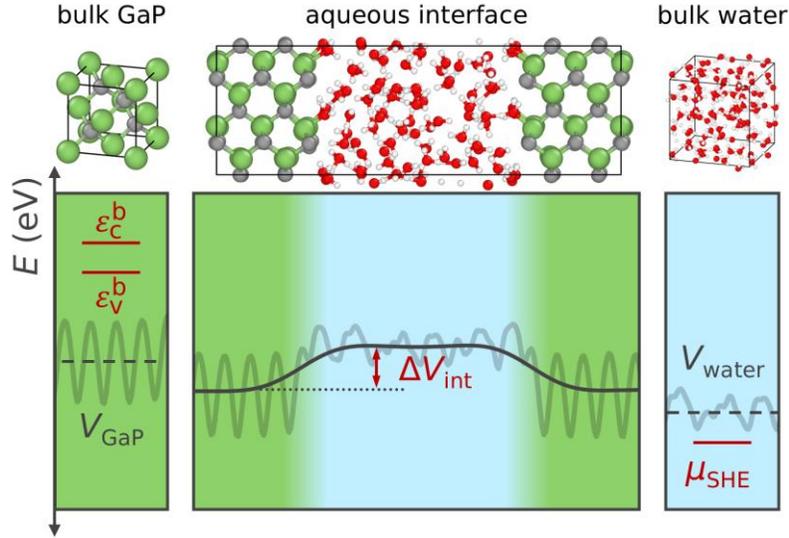

**Figure 1.** Schematic representation of the procedure followed for the determination of $\varepsilon_V^{\text{SHE}}$ and $\varepsilon_C^{\text{SHE}}$, i.e. the alignment of the semiconductor band edge positions with respect to $\mu_{\text{SHE}}$. First, the position of $\mu_{\text{SHE}}$ with respect to the averaged electrostatic potential of bulk liquid water ($V_{\text{water}}$) is calculated using thermodynamic integration and AIMD simulations. Then, the semiconductor band edges ($\varepsilon_V^b$ and $\varepsilon_C^b$) referenced to the averaged electrostatic potential of bulk GaP ($V_{\text{GaP}}$) are determined from static DFT calculations. Finally, using AIMD simulations of the GaP/water interface we calculate $\Delta V_{\text{int}}$, i.e. the electrostatic potential difference between the bulk-like regions of GaP and water. To align the band edges of GaP to $\mu_{\text{SHE}}$ eq. (1) is used.

*Computational standard hydrogen electrode potential*

The absolute standard hydrogen electrode potential corresponding to the hydrogen redox reaction, has experimentally been determined as $-4.44$ eV[21]. While this value could in principle be used for a computational alignment, it has been shown that the use of a computational potential yields much more reliable results[22–25]. Moreover, the definition of $\varepsilon_V^{SHE}$ used in this work only requires the evaluation of the SHE potential referenced to the averaged electrostatic potential of bulk liquid water ($V_{water}$). Therefore, a more computationally demanding vacuum-containing slab calculation is not required.

To evaluate $\mu_{SHE}$, we use the computational definition based on the deprotonation of a hydronium ion[22,23], according to which:

$$\mu_{SHE} = \Delta_{dp} A_{H_3O^+} + \mu_H - \Delta E_{corr} \qquad (3)$$

where $\Delta_{dp} A_{H_3O^+}$ is the deprotonation free energy of a hydronium ion in solution, $\mu_H$ is the gas-phase hydrogen chemical potential, and $\Delta E_{corr}$ is a correction term accounting for the zero-point energy of the $O - H$ bond in $H_3O^+$ and for electrostatic finite-size effects in periodic calculations.

The evaluation of $\Delta_{dp} A_{H_3O^+}$, which constitutes the most burdensome aspect of the calculation of $\mu_{SHE}$, is achieved by combining the thermodynamic integration method with AIMD simulations[26]. This calculation, the details of which can be found in the Supporting Information (SI), yields a value of 15.34 eV for $\Delta_{dp} A_{H_3O^+}$, in excellent agreement with previous estimations[23,26–28]. We use a value of 15.78 eV for $\mu_H$, evaluated as half of the density functional theory (DFT) calculated energy of a hydrogen molecule in vacuum, which is also very close to the experimentally determined value of 15.81 eV[21]. Lastly, the correction term $\Delta E_{corr}$ amounts to 0.34 eV[23]. By combing the parameters determined above, we arrive at a value of −0.78 eV for $\mu_{SHE}$.

*Bulk GaP*

Next, the bulk semiconductor band edges $\varepsilon_V^b$ and $\varepsilon_C^b$ are determined, with respect to the planar-averaged electrostatic potential, from DFT calculations. First, we optimize the structure of the GaP primitive cell using the Perdew, Burke, and Ernzerhof (PBE)[29] functional, leading to a lattice parameter of $5.51\,\text{Å}$ ($a_{exp} = 5.45\,\text{Å}$[30]). For a more accurate description of the electronic properties, we use the hybrid PBE0[31] functional, and we tune the parameter $\alpha$ that controls the fraction of the Fock exchange so that the experimental bandgap is reproduced. This method has been shown to yield very reliable results for the electronic level alignment of the aqueous interfaces of a range of semiconductors[16]. By setting $\alpha = 13$ we acquire a bandgap of 2.33 eV at 0 K and we adjust the band edges by moving both the VBM and the CBM by 0.04 eV inwards, to achieve the narrowing of the bandgap by 0.08 eV, which is observed in GaP when moving from 0 K to 300 K[32]. This procedure allows the determination of $\varepsilon_V^b$ and $\varepsilon_C^b$ at room temperature, which will be used for calculating $\varepsilon_V^{SHE}$ and $\varepsilon_C^{SHE}$ using eq. (1) and (2).

*Structural models*

Having determined $\mu_{SHE}$ with respect to the average electrostatic potential of liquid water and $\varepsilon_V^b$ and $\varepsilon_C^b$ with the corresponding level in GaP, we can now proceed with calculating $\Delta V_{int}$, the final term necessary for the band alignment, from the simulations of the solid/liquid heterostructures. For that, appropriate interface models need to be carefully chosen, considering that many different surface reconstructions have been observed for this material[33].

We start from the non-polar (110) GaP surface, which is shown in Fig. 2 a. The band alignment between this surface and water was previously studied by Guo et al[16]. Here, we repeat these simulations both to establish a reference and to provide further insight into the observed behavior. We model this surface using a 9-layer $3 \times 2$ slab with an area of $11.68 \times 11.01\,\text{Å}^2$.

The (001) facets, on the other hand, are polar, i.e. only one of the two kinds of atoms are exposed on the surface. As polar surfaces are generally electrostatically unstable, these terminations tend to reconstruct[34]. The structure of the reconstructed surfaces depends on the fabrication method and environmental conditions. Under Ga-rich conditions, the $(2 \times 4)$ mixed dimer reconstruction depicted in Fig. 2 b is the most commonly observed surface termination and will therefore be investigated[10,35]. Additionally, Jeon et al.[36] proposed a Ga-rich termination featuring $Ga - O - Ga$ bridges (Fig. S2), representing the oxidation that often takes place on the surface of GaP due to its interaction with water. As such a termination would be very relevant for water splitting applications, we considered including it in our study. However, within femtoseconds of AIMD simulations of the interface between this termination and water, deprotonation of liquid water and simultaneous protonation of the oxide surface is observed (see Fig. S2 for details). Similar observations are reported in ref. [17]. To maintain charge neutrality in the electrolyte, we therefore omit this configuration from our band alignment simulations and adopt a hydroxylated surface as a starting point instead, with adsorption of OH at bridge and atop Ga positions at 1.5ML coverage, as proposed by Wood et al[37] (Fig. 2 c).

Regarding the P-rich case, in environments lacking hydrogen, e.g. during deposition by molecular beam epitaxy, a reconstruction consisting of surface P-dimers can potentially form[10,11]. However, similar to the case of the oxide surface, within $< 0.5$ ps of AIMD simulations of the interface between this P-rich surface and liquid water, water deprotonation and hydrogen adsorption on the semiconductor surface is observed (Fig. S3). Instead, we will study the $(2 \times 2)$ reconstruction comprising P-dimers stabilized by hydrogen atoms, as shown in Fig. 2 d[38], which is much more relevant for applications that involve water interfaces. Unlike Ga-rich terminations, oxidation is also not prevalent in P-rich surfaces, and it will therefore not be studied[39,40]. All (001) facets are modelled using 15-layer $4 \times 4$ slabs with an area of $15.58 \times 15.58$ Å$^2$.

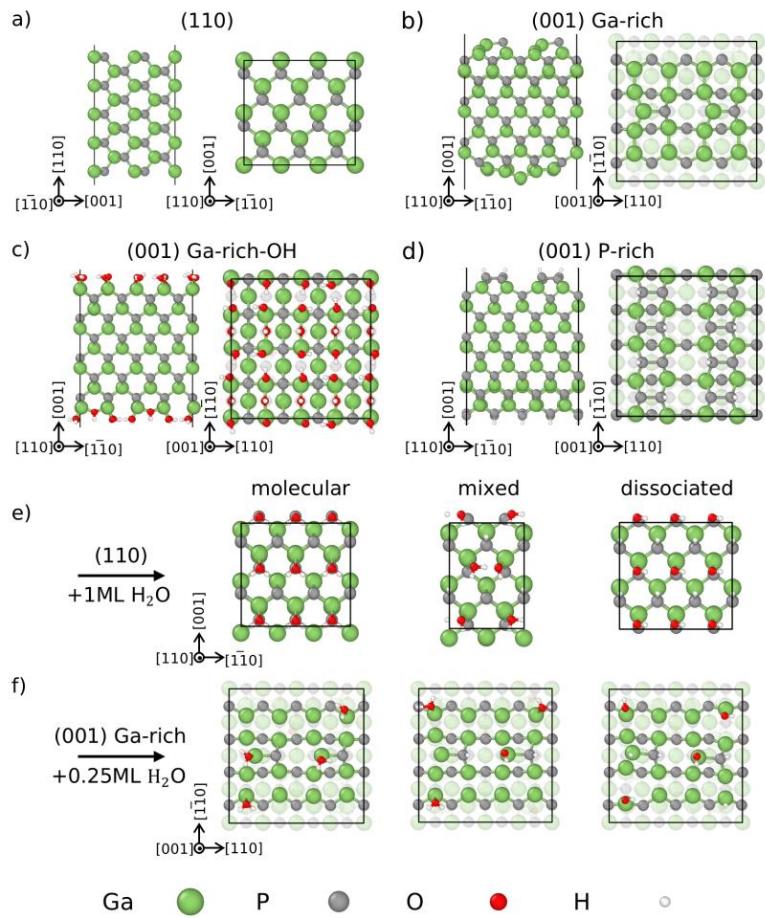

**Figure 2.** Side (left) and top (right) view of the a) (110), b) (001) Ga-rich (2 × 4) mixed dimer, c) (001) Ga-rich hydroxylated, and d) (001) P-rich (2 × 2) H-passivated P-dimer GaP surface terminations. Top view of the e) (110) and f) (001) Ga-rich (2 × 4) mixed dimer GaP surfaces with water adsorbed molecularly, in a mixed molecular-dissociated state, and dissociatively.

Another important consideration before proceeding with the construction of the solid/liquid heterostructures is the state in which water is adsorbed on the various surfaces. For the non-polar (110) termination, experimental and theoretical work points towards a mixed molecular-dissociated state as the most energetically favourable[41,42], however, AIMD simulations suggest that water is more likely to be dissociated on that surface for a wide pH range including neutral and alkaline conditions[43]. Similarly, on the (001) Ga-rich surface, experiments revealed a mixed adsorption state with a 0.25ML saturation coverage[40], but calculations showed that while dissociative water adsorption is thermodynamically more favorable, dissociation barriers are high[17]. Hence, in this work we study three different adsorption states, namely molecular, mixed, and dissociated, and compare their effect on the band alignment. In practice, we adsorb 1ML and 0.25ML of water on the (110) and the (001) terminations respectively in all three states, as shown in Fig. 2 e and f. These terminations, after structural relaxation, are used for the construction of the solid/liquid heterostructures. As will be shown below, during our simulations the composition of interfacial water remained mostly unchanged, i.e. almost no dissociation of water molecules or recombination of H/OH was observed. This can be attributed to the short time scales accessible with AIMD simulations, which do not allow the determination of the most stable interface configuration, making the explicit examination of the different scenarios explored in this study necessary. No water molecules are pre-adsorbed on the hydroxylated Ga-rich and the P-rich surfaces, as in the former case, the reactivity with water has already been considered in selecting the specific termination, while in the latter, it has been shown that no significant amounts of oxygen are found on the surface[40].

Once all surfaces have been relaxed, the interface models are created by combining the slabs with a box of water molecules, similar to the atomistic model of the interface presented in Fig. 1. The area of that box is the same as the semiconductor surface area, while the height is chosen so that the equilibrium density of liquid water at room temperature is attained. The box contains 64 molecules for the (110) surfaces (without considering the water already adsorbed on the surface, as discussed above), while 96 molecules are employed for the (001) surfaces, so that a satisfactory water layer thickness is achieved (> 10 Å), despite the larger area of those models. It should be noted that to achieve the specific adsorption pattern of the mixed molecular-dissociated state on the (110) surface, a (2 × 2) instead of a (3 × 2) cell is employed (Fig. 2 e, middle).

*Band alignment*

For band alignment calculations, 10 ps AIMD simulations of the GaP-water interface are performed, using the structures described above, which have first been equilibrated with respect to the total energy. $\Delta V_{int}$ is then calculated by averaging over time the potential difference between the bulk-like regions of water and GaP, as shown in figure S4. Using $\Delta V_{int}$ in eq. 1, the position of the semiconductor band edges is finally determined with respect to $\mu_{SHE}$, as shown in Fig. 3.

The first observation is that both the termination and how water is adsorbed on the surface have a major impact on the band alignment. In the case of the non-polar (110) surface there is a difference of more than 1 eV in the position of the band edges between molecular and dissociative adsorption (Fig. 2e), with the VBM in the former case being almost 0.5 eV above $\mu_{SHE}$. On the other hand, dissociative adsorption leads to an alignment more in line with the one often mentioned in the literature, with the VBM ~1 eV below $\mu_{SHE}$. These results are in excellent agreement with the work of Guo et al.[16], who, based on these observations, argued that dissociative adsorption is more likely on that surface, as it provides a band alignment that is closer to expectation, compared to molecular adsorption. However, during our simulations, we observe the desorption of one of the hydrogen atoms adsorbed on the (110) surface, after ~7 ps of simulation, which after migrating through the water box via the Grotthuss mechanism[44], it eventually recombines with a surface OH (see Fig. S5 for details). Note that after this desorption occurs, the potential offset starts to decrease on average, therefore we do not consider values after this event when calculating $\Delta V_{int}$ for this case (see Fig. S4). The observed desorption is an indication that, as experiments also suggest[41], a mixed molecular-dissociative adsorption state is potentially more favorable. Interestingly, the mixed state we investigate here leads to an alignment closer to the one observed for molecular adsorption, with the VBM being slightly above $\mu_{SHE}$.

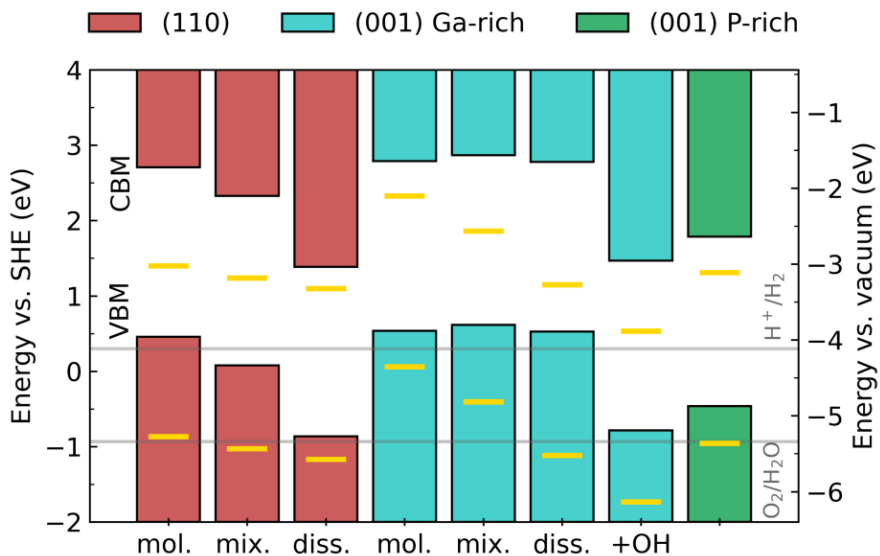

**Figure 3** Electronic level alignment of the GaP band edge positions for the various terminations with respect to $\mu_{SHE}$. The yellow lines represent the band edge positions of the corresponding terminations, as presented in Fig. 2, with respect to the vacuum level calculated from static DFT simulations of slab structures, including adsorbates, but without liquid water). The y-axes are aligned to each other assuming the experimental value of -4.44 eV for $\mu_{SHE}$. The $H^+/H_2$ redox level is plotted 0.3 eV above $\mu_{SHE}$, which is the expected value at the PZC of GaP. This value is obtained by combining the Nernst equation with a value of 4.9 for the pH at the PZC, as proposed in ref. [16] for GaP (110). The $O_2/H_2O$ redox level is plotted 1.23 eV below the $H^+/H_2$ level.

Regarding the polar (001) surface, the state in which water is adsorbed on the Ga-rich termination (Fig. 2f) does not seem to have such a significant impact on the band alignment. In all three cases (molecular, mixed, and dissociated) the VBM is positioned ~0.5 eV above $\mu_{SHE}$, very similar to the (110) case with water adsorbed molecularly. The exception to this is the fully hydroxylated Ga-rich surface (Fig. 2c) which leads to a very different alignment, with the VBM positioned almost ~1 eV below $\mu_{SHE}$, as in the (110) case with dissociative adsorption. Similar alignment is also observed for the P-rich case (Fig. 2d), where the VBM is positioned ~0.5 eV below $\mu_{SHE}$.

To gain some understanding on the behavior of the various GaP terminations, the charge distribution on the surface of the semiconductor is studied. The formation of an interface between the semiconductor and water is accompanied by the creation of bonds and, ultimately, a charge redistribution at the interface. This can result in a potential step that, depending on the direction of charge transfer, can shift the band edges to higher or lower energies[45]. To visualize this, the charge difference $\Delta\rho$ can be used, shown in Fig. 4, defined as:

$$\Delta\rho = \rho_{int} - \rho_{GaP} - \rho_w \tag{4}$$

where $\rho_{int}$ is the plane- and time-averaged electron density profile of the solid/liquid heterostructure along the z-axis (i.e. perpendicular to the interface), while $\rho_{GaP}$ and $\rho_w$ are the same profiles for the GaP and water structures (including pre-adsorbed O-/H-based species in the latter case) respectively, in the same geometry as the solid/liquid heterostructure.

Starting from the (110) surface, when water is adsorbed molecularly, a relatively flat charge density profile is obtained, with a small electron accumulation region right above the GaP surface, followed by a small depletion around 2 Å above the surface (Fig. 4 a). These two regions correspond approximately to the O and H atoms respectively, of the first layer of adsorbed water molecules. By integrating the curve, we find an overall transfer of $0.2 \times 10^{-4}$ e/Å² from water to the semiconductor surface (see Table 1). On the other hand, mixed molecular-dissociated adsorption leads to a more pronounced charge depletion at the semiconductor surface (Fig. 4 b), resulting in an electron transfer of $1.0 \times 10^{-4}$ e/Å², but this time from the semiconductor to the liquid. Dissociative adsorption shows a similar behavior as the mixed case, but with an even more severe electron accumulation around 1 Å above the interface (Fig. 4 c) and an overall transfer of $1.8 \times 10^{-4}$ e/Å² from GaP to water.

The particular charge distribution in the mixed and dissociated cases hinders the extraction of electrons from the semiconductor surface and thus increases the ionization potential, or, in other words, pushes the band edges to lower energies. The fact that in the dissociated case the charge transfer not only follows a different direction but is also much more severe than in the molecular case (almost ten times larger), is partly the reason behind the very different band alignments obtained for the two surfaces, with the energy levels in the dissociative case being more than 1 eV lower than the molecular case.

Moving to the charge difference profiles of the (001) Ga-rich surface, here we observe a small charge transfer that progresses from an electron accumulation at the subsurface, when water is molecularly adsorbed, to an electron depletion for dissociative adsorption (Fig. 4 d-f). The position of the corresponding peaks being at $z < 0$ has to do with the morphology of this termination, i.e. water interacts not only with the outermost mixed Ga-P dimers on the surface, but also with the Ga-rich layer underneath (Fig. 2b). Despite the differences in the charge density difference profiles, the band alignment for these three Ga-rich terminations yields very similar results. Lastly, the charge density difference profiles of the (001) Ga-rich hydroxylated (Fig. 4 g) and the P-rich (Fig. 4 h) terminations are qualitatively very similar to each other and to the (110) dissociated and mixed cases, with a substantial electron depletion at the surface and accumulation above it.

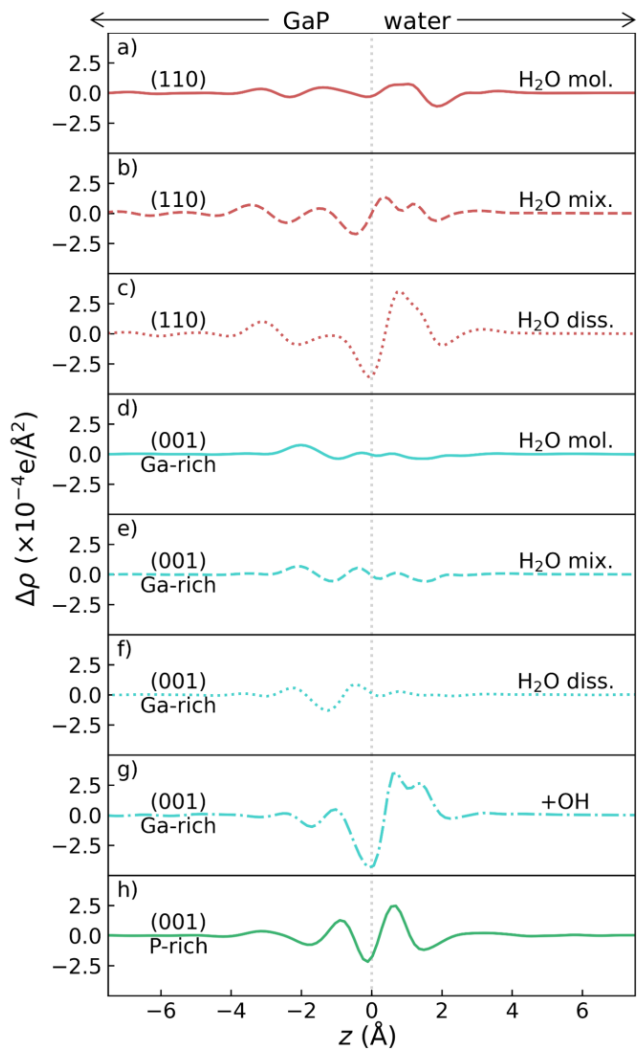

**Figure 4.** Charge difference $\Delta\rho$ given from equation (4) using charge densities of 10 ps AIMD simulations separated by 0.05 ps. Positive and negative values indicate electron accumulation and depletion respectively.

The overall charge redistribution is the largest for the Ga-rich hydroxylated termination, followed by the (110) mixed and dissociated surfaces and, lastly, the P-rich surface. While these four terminations exhibit similar charge density profiles, only three of them lead to comparable band alignments, whereas the alignment in the mixed case resembles more the terminations where the interfacial electron transfer is smaller and/or follows a different direction. This cannot be explained based on charge transfer arguments alone, therefore, another thing is at play here, which has to do with the orientation of the water molecules adsorbed on the surface of GaP[45,46]. Focusing on the (110) terminations, when 1ML of water molecules are adsorbed on a GaP surface in vacuum, the water dipoles are orientated with the H atoms slightly closer to the surface than O, and the dipole vectors form an angle of ~10° with the plane parallel to the surface (Fig. 5 a). However, when GaP is interfaced with liquid water, the surface adsorbed molecules interact with the next water layers, mainly through hydrogen bonds. This changes the surface dipole orientation, which again forms an angle of ~10° with the plane parallel to the surface but the dipoles are now directed to the opposite direction leading to a shift of the electronic levels to higher energies (Fig. 5b). To get an idea of the size of the shift a 10⁰ degree angle of the water dipole vector could cause, the expression $\Delta V = e\mathbf{p}/(\varepsilon_0 A)$ can be used, where $e$, $\mathbf{p}$, $\varepsilon_0$, and $A$ are the electron charge, dipole vector, dielectric function, and surface area respectively. By assuming a value of 1.85 Debye for the dipole moment of a water molecule[47], we get $\Delta V \sim 0.5$ eV. This is the shift of the electronic levels caused by six water molecules on the (110) GaP surface with their water dipoles forming 10⁰ degree angles with that surface. Therefore, seemingly small rearrangements of the surface molecules can have a big impact on the interface electronic properties.

As this process is more relevant in the molecular and mixed cases, than in the dissociative case where no significant amount of water dipoles is found near the surface, this explains why the former terminations lead to much higher band edge positions than the dissociated case. This is further confirmed by the band alignment in vacuum, which is denoted by the yellow lines in Fig. 3. Specifically, in vacuum, where the water dipoles are pointed to the opposite direction than in the aqueous interface, the variation of the band edge positions for the three different adsorption cases is much smaller, with ~0.5 eV difference between molecular and dissociative adsorption and with the mixed case exhibiting intermediate behavior.

**Table 1**. Charge transfer along the GaP/water interface and average angle between the H₂O dipole vector and the plane parallel to the GaP surface.[1]

| Structure | Charge transfer (×10⁻⁴ e/Å²) | H₂O Dipole Angle (°) |
|---|---|---|
| (110) H$_2$O mol. | +0.2 | -12 |
| (110) H$_2$O mix. | -1.0 | -31 |
| (110) H$_2$O dis. | -1.8 | - |
| (001) Ga-rich H$_2$O mol. | +0.5 | -25 |
| (001) Ga-rich H$_2$O mix. | +0.4 | -21 |
| (001) Ga-rich H$_2$O dis. | -0.1 | -38 |
| (001) Ga-rich +OH | -2.5 | -41 |
| (001) P-rich | -0.6 | -54 |

The orientation of the surface water dipoles can also explain why the Ga-rich (001) terminations (excluding the hydroxylated one) exhibit similar band alignments, despite the differences in their charge density difference profiles. Again, we hypothesize that this is due to the water dipoles on the surface, with average angles reaching up to 38⁰ as can be seen in Table 1, accompanied by an increasing number of water molecules close to the surface from the molecular, to the mixed, to the dissociated case, as can be seen in Fig. S6. In contrast, in vacuum, the observed trend is more in line with what one would expect by looking at the charge density difference profiles alone. It is therefore interesting to also note that while some (although not all) trends regarding the band edge positions are accurately captured with simulations of slabs in vacuum, our results indicate that taking the effect of solvent into account is necessary for accurate predictions of the interface band alignment.

---

[1] The charge transfer is calculated by integrating the charge difference profiles of Fig. 4 from the middle of the slab to the interface and from the interface to the middle of the liquid water box and averaging the absolute values. Positive values indicate electron transfer from water to the semiconductor. The dipole angles are averaged for all water molecules with an O atom within 3 Å from a P or Ga surface atom and for timeframes separated by 0.05 ps for the same time intervals that are used for the averaging of $\Delta V_{\text{int}}$ as shown in Fig. S4. Negative values indicate that the dipoles point towards the surface.

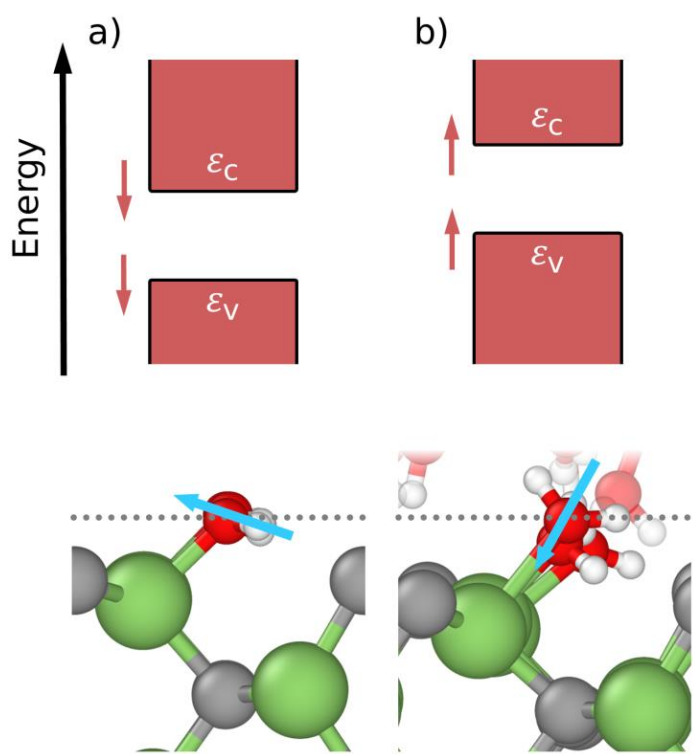

**Figure 5.** Schematic representation of the band edge position shift when water molecules are adsorbed a) with the dipole vector pointing out of the surface as is the case for the (110) surface in vacuum or b) with the dipole vector pointing towards the surface as is the case on the same surface in contact with liquid water.

Overall, the (110) dissociated, (001) Ga-rich hydroxylated, and (001) P-rich terminations are the only three terminations with a reasonably good alignment with the levels associated with the HER and the OER. Specifically, The CBM of these terminations lies ~1.0-1.5 eV above $H^+/H_2$ level, a prerequisite for the use of GaP as a photocatalyst in HER, as can be seen in Fig. 3. For all other terminations, the gap between their CBM and this level is much larger (~2.5-3.0 eV). In addition, in the (110) dissociated and the (001) Ga-rich hydroxylated terminations the VBM lies very close to the $O_2/H_2O$ level, indicating that these terminations could also potentially be used for the oxidation of water.

From the above analysis it becomes clear that the surface termination of GaP can have a huge impact on the ability of the material to catalyze the reactions associated with water splitting. Overall, it seems that a termination that promotes electron transfer from the surface towards the electrolyte leads to a band alignment much more favorable for water splitting than configurations with no significant charge displacement. Although these simulations are conducted at the point of zero charge (PZC), a different pH should not directly lead to a different alignment, if we assume both the HER and the OER levels and the band edges to follow a Nernstian behavior[6,43,48]. However, the composition of water on the surface would be significantly impacted by the pH, which could in turn favor or compromise the band alignment and eventually the efficacy of GaP for solar-driven water splitting.

More specifically, in the non-polar (110) surface it has been shown that in neutral conditions (5 < pH < 9), water is expected to be mostly dissociatively adsorbed, while molecular adsorption becomes more prevalent in acidic conditions. The concentration of adsorbed protons is also expected to drop in very alkaline conditions[43]. This means that the alignment of the (110) surface is expected to be more favorable in neutral to moderate alkaline conditions. On the other hand, to mitigate the unfavorable alignment owing to the molecular adsorption of water in neutral to acidic conditions, the surface could be passivated with electron accepting species, which could promote a charge distribution at the interface that can lead to the desired band alignment. The pH-dependent water surface coverage of the (001) terminations is not known, as to our knowledge no similar studies have been conducted for those surfaces. However, a similar approach would also work in this case, namely passivating the surface with electron accepting species.

Besides the concentration and dissociation degree of interfacial water, the orientation of water molecules in different conditions can also vary significantly. However, this property could also be controlled, as has been shown to be the case for Pt surfaces, where alkalis can be used to promote a water molecule orientation with the hydrogens pointing towards the surface[49]. Such processing would not only be beneficial for the band alignment in the case of GaP, but could also promote reaction kinetics, as interaction between the semiconductor surface and H can facilitate the water splitting reactions[50].

*Interface structure properties*

Catalytic efficiency does not only depend on the electronic properties, but also on the structure of the surface of the semiconductor as well as on the interfacial water structure. Although many studies have attempted to elucidate the relationship between these properties, the ideal interface structure remains unclear[51]. However, there are still conclusions to be drawn regarding the expected performance of the various terminations studied here, by looking at their aqueous interfaces.

In Fig. 6, the GaP/water interface structure for the three terminations that exhibit favorable electronic level alignment for water splitting, namely the (110) termination with dissociatively adsorbed water, the (001) Ga-rich hydroxylated termination, and the (001) P-rich termination are compared. One of the most striking observations emerging from our simulations is the relatively limited interaction between the latter termination and liquid water. Specifically, there is a noticeable gap between the P-rich surface and water, with the H-passivated P dimers exhibiting hydrophobic behavior. This is also attested from the very low number of O atoms within 3 Å from a Ga or P atom, compared to the other surfaces, as shown in Fig. 6 and S6. This agrees with experimental observations of very low water coverage of the P-rich surface[40]. Similar behavior is also observed in the (110) termination, where water does not interact with the P-H sites, however, as the surface hydroxyls form hydrogen bonds with the water molecules, the latter form bridge-like arrangements on the surface. The (001) Ga-rich termination looks completely different. Specifically, both for the non-polar and for the Ga-rich surface there is approximately one H-bond per adsorbed -OH, between the surface and the liquid water. However, the higher -OH coverage of the Ga-rich surface ensures a much stronger interaction between that termination and water.

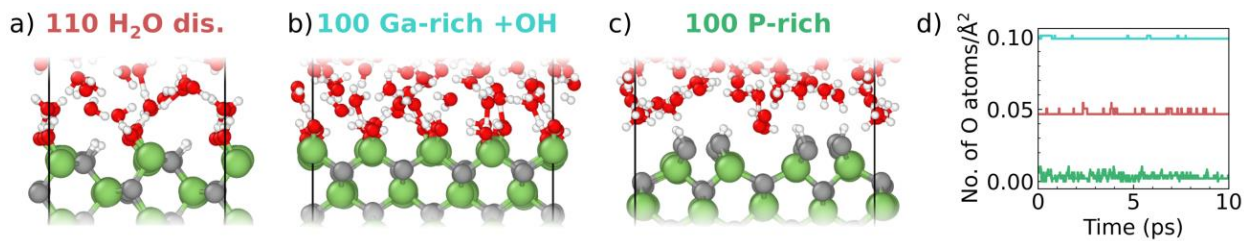

**Figure 6**. Atomistic representation of the GaP/liquid water interface for a) the (110) termination with water adsorbed dissociatively, b) the (001) Ga-rich hydroxylated termination, and c) the (001) P-rich termination. d) Number of O-containing species per surface area within 3 Å of a P or Ga atom for the three different GaP terminations over 10 ps of AIMD simulations.

As interaction between water and the photocatalyst surface is a prerequisite for efficient water splitting, according to the above the (001) P-rich termination is clearly not the best choice for the desired application. In contrast, the non-polar (110) and the (001) Ga-rich hydroxylated terminations with their interfacial hydrogen bond networks that form due to the surface hydroxyl groups would be much better candidates for efficient water splitting. Indeed, several examples exist in the literature, where surface hydroxyl groups have been employed to increase the hydrophilicity of surfaces, thus enhancing water adsorption and ultimately increasing HER performance[52]. However, the degree of hydroxylation of the surface depends on a lot of factors, such as fabrication method, as well as environmental conditions, like pH. We saw for instance that in neutral conditions, on the non-polar GaP surface a mixed molecular-dissociated adsorption of water might be more favorable, which would negatively impact the band alignment, besides modifying the hydrogen bond network. That leaves us with the Ga-rich hydroxylated surface, which however might be unattainable under certain conditions due to high energy barriers for the dissociation of water on that surface[17], but these can still potentially be overcome with appropriate control of the fabrication conditions[53]. Another caveat is that this surface is known to be prone to corrosion[15], however, surface hydroxylation could act as passivation against it[37]. Overall, the Ga-rich hydroxylated surface exhibits the most promising properties among the terminations studied here, with a very favorable electronic level alignment and an interface structure with properties desirable for photocatalysis. In any case, the stability and morphology of this interface under different conditions needs to be further explored to ensure optimal performance during operation of a PEC cell.

CONCLUSION

In this work we have investigated the aqueous interfaces of various polar and non-polar GaP terminations using AIMD simulations. By aligning the electronic levels of GaP to the computational standard hydrogen electrode (SHE) potential, we demonstrate that both surface termination and the structure of interfacial water profoundly affect band edge positions. Specifically, electron transfer from the GaP surface to interfacial water yields a favorable alignment of electronic levels for water-splitting applications. In contrast, electron transfer in the opposite direction shifts the band edges to energies that are too high for efficiently catalyze the HER. In addition, the orientation of surface water dipoles is found to significantly influence band edge positions. When the dipole vectors of adsorbed water molecules point towards the surface, the resulting electronic alignment shifts away from the optimal range of catalytic activity. Importantly, not all terminations that exhibit favorable electronic level alignment also possess interfacial structural properties conductive to efficient reaction kinetics. Among the terminations studied, the (001) Ga-rich hydroxylated surface emerges as the most promising candidate for photoelectrochemical applications. Nonetheless, a deeper understanding of the interfacial composition under realistic operating conditions is essential to evaluate catalytic effectiveness of the different GaP terminations in water splitting reactions.

METHODS

Ab initio molecular dynamics simulations of liquid water for the determination of $\mu_{\text{SHE}}$ and of the solid/liquid interface for the electronic level alignment were performed using the freely available CP2K/Quickstep package[54]. Analytic Goedecker–Teter–Hutter (GTH) pseudopotentials were used to represent the core electrons[55]. A triple-ζ correlation-consistent polarized basis set (cc-pVTZ) was used for O, H, and P, and a double-ζ (cc-pVDZ) for Ga[56]. The plane wave cutoff for the electron density was set to 600 Ry. The Brillouin zone was sampled only at the Γ-point. We employed the PBE functional[29], along with the rVV10 functional to account for nonlocal van der Waals interactions[57]. Following ref. [58] we set the b parameter of the rVV10 functional to the value of 9.3, as it was found to best reproduce the density and structural properties of liquid water. All molecular dynamics simulations were conducted in the NVT ensemble, and the temperature was set to 350K. We used a Nosé–Hoover thermostat to control the temperature[59–61], while a timestep of 0.5 fs was employed.

The PBE exchange–correlation functional within the generalized gradient approximation (GGA)[29] was used for the structural relaxation of the unit cell of GaP with CP2K/Quickstep. The rVV10 functional with the b parameter set to 9.3 was also included for the structural relaxation of the various GaP slab models. For the determination of the band edge positions with the hybrid PBE0[31], the Vienna Ab Initio Simulation Package (VASP)[62–64] was employed, along with a plane wave cutoff of 400 eV and a Γ-centered 6x6x6 k-point grid.

AUTHOR INFORMATION
**Corresponding Authors**

* laurent.pedesseau@insa-rennes.fr

*charles.cornet@insa-rennes.fr

ACKNOWLEDGMENT

This research was supported by the Brittany Region under the "Stratégie d'Attractivité Durable (SAD)" funding and by the "France 2030" program of the French National Research Agency, NAUTILUS Project (Grant no. ANR-22-PEHY-0013). DFT calculations were performed at Institut FOTON, and the work was granted access to the HPC resources of TGCC/CINES under the allocation 2024-A0160911434 and 2024-A0180911434 made by GENCI.